\documentclass[10pt]{revtex4}
\usepackage{amssymb}
\usepackage{latexsym}
\usepackage{epsfig}
\usepackage{amsmath}
\usepackage{color}

\begin{document}

\title{A Generalization to the Rastall Theory and Cosmic Eras}
\author{H. Moradpour$^1$\footnote{Corresponding Author: h.moradpour@riaam.ac.ir}, Y. Heydarzade$^{2}$\footnote{heydarzade@azaruniv.edu}, F. Darabi$^{2}$\footnote{f.darabi@azaruniv.edu},
Ines G. Salako$^{3,4,5}$\footnote{inessalako@gmail.com}}
\address{$^1$ Research Institute for Astronomy and Astrophysics of
Maragha (RIAAM), Maragha 55134-441, Iran\\
$^2$ Department of Physics, Azarbaijan Shahid Madani University,
Tabriz, 53714-161, Iran\\
$^3$ Institut de Math\'ematiques et de Sciences Physiques
(IMSP), Universit\'e de Porto-Novo, 01 BP 613 Porto-Novo, B\'enin\\
$^4$ D\'epartement de Physique, Universit\'e d'Agriculture de
K\'etou, BP 13 K\'etou,  B\'enin\\
$^5$ African Institute for Mathematical Sciences(AIMS), $6$
Melrose Road, Muizenberg, $7945$, South Africa}
\begin{abstract}
A generalized version for the Rastall theory is proposed showing
the agreement with the cosmic accelerating expansion. In this
regard, a coupling between geometry and the pressureless matter
fields is derived which may play the role of dark energy
responsible for the current accelerating expansion phase.
Moreover, our study also shows that the radiation field may not be
coupled to the geometry in a non-minimal way which represents that
the ordinary energy-momentum conservation law is respected by the
radiation source. It is also shown that the primary inflationary
era may be justified by the ability of the geometry to couple to
the energy-momentum source in an empty flat FRW universe. In fact,
this ability is independent of the existence of the
energy-momentum source and may compel the empty flat FRW universe
to expand exponentially. Finally, we consider a flat FRW universe
field by a spatially homogeneous scalar field evolving in
potential $\mathcal{V}(\phi)$, and study the results of applying
the slow-roll approximation to the system which may lead to an
inflationary phase for the universe expansion.
\end{abstract}

\maketitle

\section{Introduction}

The origins of the primary inflationary era \cite{inflation1,
inflation2, inflation3, inflation4}, current accelerating phase of
the universe expansion \cite{expansion1, expansion2, expansion3,
expansion4, expansion5} as well as the dark matter problem
\cite{DM1, DM2, DM3} are some of the big puzzles in the standard
model of cosmology. Our insufficient understanding of these
problems  leads the coincidence and fine-tuning problems
\cite{roos, coin, fine1, fine2}. In order to solve the above
mentioned problems, some authors have been introduced a new type
of energy-momentum source \cite{Rev3,Rev2,Rev1,mod}. In another
approach, physicist try to solve the above problems by modifying
the Einstein field equations
\cite{lobo,meeq,rastall,cmc,cmc1,cmc2}. In this line, one may
refer to the scalar-tensor gravity \cite{Faraoni}, vector-tensor
theories \cite{vector}, tensor-vector-scalar theories \cite{tvs},
quadratic gravity \cite{quad}, Chern-Simons theories
\cite{chern1}, massive gravity \cite{massive1, massive2} and
Gauss-Bonnet theory \cite{gauss}, for a review see also
\cite{LRL}. Scalar-tensor (ST) theories of gravity are the
simplest alternative to Einstein's general theory of gravity (GR)
and  have a long history. The first attempts are done by Jordan
\cite{ST1,ST10}, Fierz \cite{ST2}, and Brans-Dicke \cite{ST3}.
These theories possess just one massless scalar field and with a
constant coupling strength to matter fields. These works were
generalized later to the theories in which the scalar field has a
dynamic coupling to the matter fields and/or an arbitrary
self-interaction in \cite{ST4, ST5, ST6} as well as to the theory
with multiple scalar fields  \cite{ST7}. In the vector-tensor
theories of gravity, in addition to the metric tensor,  the
gravitational action is modified by adding a vector field that is
non-minimally coupled to gravity. Studying these theories refer to
the works by Will, Nordtvedt and Hellings \cite{VT1, VT2, VT3},
see also \cite{VT4, VT5}. The tensor-vector-scalar  theory is
proposed by Bekenstein \cite{Beken} where the standard Einstein
tensor field of General Relativity (GR) is coupled to a vector
field as well as a scalar field, hence the theory is called by
this name. This theory is a relativistic  version of Modified
Newtonian Dynamics (MOND) \cite{mond} reproducing MOND in the weak
field limit. The most important advantage to adopt
tensor-vector-scalar  theory refers to the explanation of many
galactic and cosmological observations without the need for dark
matter \cite{mond1, mond2}. The quadratic gravity theories are
based on the idea of adding appropriate quadratic terms in the
Riemann and Ricci tensors or the Ricci scalar inspired by the
string or quantum gravity theories \cite{quadrat}. Chern-�Simons
gravity is the special case of the quadratic theories including
only the parity-violating term
$^{*}RR={{^{*}R^{\alpha}}_{\beta}}^{\gamma\delta}{R^{\beta}}_{\alpha\delta\gamma}$
in which
${{^{*}R^{\alpha}}_{\beta}}^{\gamma\delta}=\frac{1}{2}\epsilon^{\gamma\delta\rho\sigma}{R^{\alpha}}_{\beta\rho\sigma}$
\cite{chern2}. Massive gravity theories are new attempts which
attribute a mass to the putative �graviton. The simplest work in
this line and in a ghost-free manner suffers from the  van
Dam-Veltman-Zakharov (vDVZ) discontinuity problem \cite{mass1,
mass2}. Due to the three additional helicity states for the
massive spin-2 graviton, the limit of small graviton mass does not
coincide with the Einstein GR. As an instance, the predicted
perihelion advance violates the previous observational
experiments. In order to resolve the vDVZ problem, a new model was
introduced  by Visser by considering a non-dynamical flat
background metric \cite{viss}. Gauss-Bonnet theory is built on
adding the quadratic combination of two Riemann tensor to the
Einstein-Hilbert action in which it does not increase the
differential order of the resulting equations of motion
\cite{Bonnet1, Bonnet2}. In most of these modified theories, the
energy-momentum source is described by a divergence-free tensor
which couples to the geometry in a minimal way \cite{lobo,meeq}.
However, it is worthwhile mentioning that this property of the
energy-momentum tensor, which leads to the energy-momentum
conservation law, is not obeyed by the particle production process
\cite{motiv1,motiv2,motiv20,motiv3,motiv4}. Therefore, it is not
unreasonable to consider a non-divergence-free energy-momentum
tensor and look for a new gravitational theory. In this regards,
P. Rastall firstly considered such kind of sources and introduced
a modification to the Einstein field equations \cite{rastall}.
Also, there is another theory known as the curvature-matter theory
of gravity \cite{cmc,cmc1,cmc2}, in which, similar to the Rastall
theory, the matter and geometry are coupled to each other in a
non-minimal way meaning that the ordinary energy-momentum
conservation law is not valid. However, it is important to stress
that all of the potential alternatives to the general theory of
relativity must be viable. This means that they must be metric
theories in order to be in agreement with the Einstein equivalence
principle, which is today supported by a very strong empirical
evidence, and that they must pass the solar system tests
\cite{LRL}. On the other hand,  the recent starting of the
gravitational wave (GW) astronomy with the event GW150914, that is
the first historical detection of GWs \cite{GW1} could be
fundamental for discriminating about various modified theories of
gravity because some differences among such theories can be
emphasized in the linearized theory of gravity and, in principle,
can be found by GW experiments, see \cite{GW2, GW3} for details.

In this work, we proposed a generalized Rastall theory to show that
a coupling between the geometry and matter fields helps us in
providing an geometric interpretation for the dark energy and thus
the current accelerating expansion phase of the universe. The main
point in favor of the Rastall theory and its generalized version is
that the usual conservation law on $T_{\mu\nu}$ is tested only in
the flat Minkowski space-time or specifically in a gravitational
weak field limit. Indeed, this theory reproduces a phenomenological
way for distinguishing features of quantum effects in gravitational
systems, i.e the violation of the classical conservation laws
\cite{motiv4, cmc, conserv2}. Also, one may find that the condition
${T^{\mu\nu}}_{;\mu}\neq0$ is phenomenologically confirmed by the
particle  creation process in cosmology \cite{motiv1, motiv2,
motiv20, motiv3, prd, particle5, Calogero1, Calogero2, Velten}. One
also may refer to \cite{neutrast} in favor of the viability of the
original Rastall theory and our proposed generalization. In this
work, it is shown that the restrictions on the Rastall parameter are
of the order of $\leq1\%$ with respect to the corresponding value of
the general theory of relativity. In other words, the results in
\cite{neutrast}  are a confirmation that the Rastall theory and its
generalization are viable theories, in the sense that the deviation
of any extended theory of gravity from the standard general theory
of relativity must be weak. Beside the current accelerating
expansion phase of the universe, the radiation dominated era in this
framework is also addressed. Moreover, we will show that the ability
of the geometry to couple with the energy-momentum source may
produce the primary inflationary era in our generalized version of
the Rastall theory.

The paper is organized as follows. After reviewing the original
Rastall theory in the next section, we address a generalization to
this theory in the third section. Section ($\textmd{IV}$) includes
some general remarks on the constructed new theory in FRW
universe. In section ($\textmd{V}$), considering a flat FRW
universe filled by a pressureless matter, we show that a
non-minimal coupling between the geometry and the energy-momentum
source may be considered as an origin for the dark energy and thus
the current accelerated phase of the universe expansion. In
section ($\textmd{VI}$), the radiation dominated era in our
generalization of the Rastall theory is investigated. In section
($\textmd{VII}$), we study two methods to model the primary
inflationary era of the universe in our formalism. Finally,
section ($\textmd{VIII}$) is devoted to the summary and concluding
remarks.

\section{A brief review on the Rastall theory}
Based on the Rastall theory, the ordinary energy-momentum
conservation law is not always available in the curved spacetime
and therefore we should have \cite{rastall}
\begin{eqnarray}\label{rastal}
T^{\mu \nu}_{\ \ ;\mu}=\lambda^{\prime} R^{;\nu},
\end{eqnarray}
where $R$ and $\lambda^{\prime}$ are the Ricci scalar of the
spacetime and the Rastall constant parameter, respectively. In
fact, $\lambda^{\prime}$ is a measure of the tendency of the
geometry (matter fields) to couple with the matter fields
(geometry) leading to the changes into the matter fields
(geometry). This equation leads to the following field equations
\begin{eqnarray}\label{r1}
G_{\mu \nu}+\kappa^{\prime}\lambda^{\prime} g_{\mu
\nu}R=\kappa^{\prime} T_{\mu \nu},
\end{eqnarray}
which can finally be rewritten as
\begin{equation}\label{ein}
G_{\mu \nu}=\kappa^{\prime} S_{\mu\nu},
\end{equation}
where $\kappa^{\prime}$ is the Rastall gravitational coupling
constant and $S_{\mu\nu}$ is the effective energy-momentum tensor
defined as
\begin{equation}\label{senergy}
S_{\mu\nu}=T_{\mu\nu}-\frac{\kappa^{\prime}\lambda^{\prime}
T}{4\kappa^{\prime}\lambda^{\prime}-1}g_{\mu\nu}.
\end{equation}
In fact, in this theory the matter fields and geometry are coupled
to each other in a non-minimal way \cite{rastall,cmc,cmc1,cmc2}
and its compatibility with some observational data have firstly
been shown by Al-Rawaf and Taha \cite{al1,al2}. Moreover, since
the particle production process during the cosmos evolution does
not respect the energy-momentum conservation law
\cite{motiv1,motiv2,motiv20,motiv3,motiv4}, the Rastall theory may
be considered as a classical background formulation for this
phenomena \cite{prd}. Finally, we should mention that
Eq.~(\ref{r1}) implies that
$R(4\kappa^{\prime}\lambda^{\prime}-1)=\kappa^{\prime} T$ where
$T$ is the trace of energy-momentum tensor. Therefore, because
$\lambda^{\prime}$ is constant and the
$R(4\kappa^{\prime}\lambda^{\prime}-1)=\kappa^{\prime} T$
condition applies to all spacetimes and energy-momentum sources,
the $\kappa^{\prime}\lambda^{\prime}=\frac{1}{4}$ case is not
allowed \cite{rastall}. More studies on the various aspects of
this theory can be found in
\cite{smal,neutrast,obs1,obs2,rastbr,rastsc,cosmos3,rascos1,rasch,more1,more2,more3,more4,hm,msal}.

\section{The generalized Rastall theory with varying Rastall parameter}

Basically, Rastall assumed that for all of the spacetimes and
energy-momentum sources, the ratio of the flow of the
energy-momentum tensor ($T^{\nu\mu}_{\ \ \ ;\mu}$) to the Ricci
scalar divergence ($R^{;\nu}$) is constant ($\lambda^{\prime}$).
It means that the evolutions of energy-momentum source and also
the geometry do not affect this ratio. As an example, consider the
matter dominated era in the universe history. The energy density
of matter decreases during the universe expansion, but the
mentioned ratio is a constant parameter in Rastall theory
\cite{al1,al2} meaning that the coupling between energy-momentum
source and geometry is constant, and is not affected by the
evolution of the cosmic system. In fact, it is a very restricting
condition to assume the evolution of system does not affect the
mutual coupling. In addition, since the mutual coupling is a
constant parameter in Rastall gravity, it did not continuously
change during the universe evolution \cite{al1,al2,prd}. Indeed,
since the cosmic evolution is a continues process \cite{roos}, it
is a reasonable expectation that the mutual coupling between the
energy-momentum sources and the geometry should be varying
gradually and smoothly. Therefore, at least theoretically, it is
not prohibited to generalize the Rastall theory as
\begin{eqnarray}\label{gr0}
T^{\mu\nu}_{\ \ \ ;\mu}=(\lambda R)^{;\nu},
\end{eqnarray}
leading to
\begin{eqnarray}\label{bian}
(T_{\mu \nu}-g_{\mu\nu}\lambda R)^{;\nu}=0.
\end{eqnarray}
Now, regarding the Bianchi identity, i.e $G_{\mu\nu}^{\ \ \
;\nu}=0$, we obtain
\begin{eqnarray}
G_{\mu\nu}=\kappa(T_{\mu \nu}-\lambda g_{\mu\nu}R),
\end{eqnarray}
where $\kappa$ is a constant and finally, we reach at
\begin{eqnarray}\label{gr}
G_{\mu\nu}+\kappa\lambda g_{\mu\nu}R=\kappa T_{\mu \nu}.
\end{eqnarray}
Although this result looks like to the field equations of the
original Rastall theory (\ref{r1}), here, $\lambda$ is not
generally constant. Just the same as $\lambda^{\prime}$ in the
Rastall theory, $\lambda$ is a measure for the strongness of the
coupling between the geometry to the matter fields. As it is
apparent, the Einstein field equations are recovered in the
appropriate limit of $\lambda=0$, a limit in which the matter
fields and geometry are coupled to each other in a minimal way.
\section{FRW metric and general remarks on the mutual non-minimal coupling between the geometry and matter fields}

The line element of the FRW universe is written as
\begin{eqnarray}\label{rw}
ds^2=-dt^2+a(t)^2[\frac{dr^2}{1-kr^2}+r^2(d\theta^2+sin(\theta)^2d\phi^2)],
\end{eqnarray}
where $a(t)$ is the scale factor and $k=-1,0,1$ is the curvature
parameter corresponding to the open, flat and closed universes,
respectively. If the universe is filled by an energy-momentum
source with $T^\mu_\nu=diag(-\rho,p,p,p)$ in which $\rho$ and $p$
are the energy density and pressure of the cosmic fluid,
respectively, then using Eq.~(\ref{gr}), the Friedmann equations
in a flat FRW universe are given as
\begin{equation}\label{fr1}
(12\kappa\lambda-3)H^2+6\kappa\lambda \dot{H}=-\kappa\rho,
\end{equation}
and
\begin{equation}\label{fr2}
(12\kappa\lambda-3)H^2+(6\kappa\lambda-2) \dot{H}=\kappa p.
\end{equation}
Here, $H=\frac{\dot{a}}{a}$ denotes the Hubble parameter, and the
dot sign indicates the derivative with respect to the cosmic time
$t$. In this manner, from Eq.~(\ref{gr0}), one easily obtains
\begin{eqnarray}\label{emc}
\frac{d(\rho+\lambda R)}{dt}+3H(\rho+p)=0,
\end{eqnarray}
meaning that the $\lambda R$ term is the energy density
corresponding to the ability of geometry to couple with the
energy-momentum sources in a non-minimal way ($\lambda\neq0$). It
is worthwhile mentioning here that for an empty spacetime where
$\rho=p=0$, we should have $\frac{d(\lambda R)}{dt}=0$. In
addition, Eq.~(\ref{emc}) can also be rewritten as
\begin{eqnarray}\label{emc1}
\rho+\rho_g=-\int3H(\rho+p)dt,
\end{eqnarray}
where $\rho_g\equiv\lambda R$. It is obvious that, in the absence
of the ability of the geometry to couple with the energy-momentum
sources in a non-minimal way ($\lambda=0$), the usual
energy-momentum conservation law and the Einstein field equations
can be recovered through the equations (\ref{gr}) and (\ref{emc}).
In the following sections, we study the role of the non-minimal
coupling between geometry and energy-momentum sources in the
various expansion phases of the flat FRW universe.
\section{Matter dominated era and an accelerating universe}

Consider a flat FRW universe with the scale factor $a$ filled by
the pressureless dust matter fields. Using the equation
(\ref{gr}), one obtains the Friedmann equations as
\begin{equation}\label{friedman1}
(12\kappa\lambda-3)H^2+6\kappa\lambda \dot{H}=-\kappa\rho_m,
\end{equation}
and
\begin{equation}\label{friedman2}
(12\kappa\lambda-3)H^2+(6\kappa\lambda-2) \dot{H}=0,
\end{equation}
where $\rho_m$ denotes the energy density. It is clear that, at
the $\lambda\rightarrow0$ limit, the equations (\ref{friedman1})
and~(\ref{friedman2}) reduce to those of the matter dominated era
in the standard cosmology \cite{roos}. In addition, Eq.~(\ref{gr})
leads to $R=-\frac{\kappa}{4\kappa\lambda-1}\rho_m$ for a dust
source requiring that we should have
$\kappa\lambda\neq\frac{1}{4}$ for $\rho_m\neq0$ in agreement with
the Rastall's original hypothesis \cite{rastall}. For the
deceleration parameter, defined as $q=-1-\frac{\dot{H}}{H^2}$
\cite{roos}, one can use Eq.~(\ref{friedman2}) to obtain
\begin{eqnarray}\label{dece1}
q(z)=1+\frac{1}{6\kappa\lambda(z)-2},
\end{eqnarray}
where $z$ denotes the redshift. It is obvious that the
deceleration parameter of the matter dominated era in the Einstein
regime ($q=\frac{1}{2}$) can be covered in the appropriate limit
of $\lambda=0$.

For a flat FRW universe filled by a pressureless matter, the
continuity equation can be written as
\begin{eqnarray}\label{cont1}
\dot{\rho}_m+3H\rho_m=\frac{d}{dt}(\frac{\kappa\lambda}{4\kappa\lambda-1}\rho_m).
\end{eqnarray}
If the pressureless source does not interact with geometry, then
this equation is decomposed into the following equations
\begin{eqnarray}\label{cont12}
\dot{\rho}_m+3H\rho_m&=&0,\nonumber\\
\frac{d}{dt}(\frac{\kappa\lambda}{4\kappa\lambda-1}\rho_m)&=&0,
\end{eqnarray}
meaning that the ordinary energy-momentum conservation law is
valid. Therefore, $\lambda=0$ is a simple solution to the
$\frac{d}{dt}(\frac{\kappa\lambda}{4\kappa\lambda-1}\rho_m)=0$
equation leading to the ordinary Einstein field equations. Now,
for a non-interacting universe, it is easy to check that
equation~(\ref{cont12}) (or equally Eq.~(\ref{cont1})) admits the
following solution
\begin{eqnarray}\label{mym}
&&\rho_m=\rho_0 a^{-3},\nonumber\\
&&\lambda(a)=\frac{1}{4\kappa+\kappa C
\rho_m}=\frac{1}{4\kappa+\kappa\alpha a^{-3}},
\end{eqnarray}
where $\rho_0$ and $C$ are integration constants and thus
$\alpha=C\rho_0$ is a constant. It is obvious that we have $\lambda=\frac{1}{4\kappa}$ 
in the absence of dust source, i.e for $\rho_m=0$. Here, we only
considered a simple situation in which there is no energy exchange
between the geometry and matter source. In this case, the
existence of matter source only affects the ability and tendency
of geometry to couple with energy source, and it does not lead to
an energy exchange between the geometry and matter source, and
thus a palpable mutual interaction between them. By the palpable
interaction, we mean an interaction leading to a visible and
measurable energy exchange between the components of system.
Therefore, it seems that the non-minimal coupling between geometry
and the matter source has some indirect, complex and non-local
aspects hidden until now, a result in line with some previous
works claiming that the probable non-local features of mutual
relation between geometry and the energy sources may be considered
as an origin for the dark sectors of the cosmos
\cite{mash,mash1,mash2}. It is also useful to note that, even in
the simplest case of~(\ref{cont12}), the properties of geometry,
including its curvature and $\lambda$, are determined by the
energy sources filling it. This is in agreement with the general
relativity backbone, where the curvature of the geometry (as its
property) is specified by the energy sources filling that. In a
more realistic case, they may exchange energy with each other, and
therefore, one cannot always decompose Eq.~(\ref{cont1}) into
Eq.~(\ref{cont12}). Now, inserting equations (\ref{my})
into~(\ref{friedman1}) and~(\ref{friedman2}), respectively, and
combining the results with each other, one reaches
\begin{eqnarray}
H(a)=H_0\sqrt{\frac{a^3+\alpha}{a^3}},
\end{eqnarray}
where $H_0=\frac{\kappa\rho_{0}}{3\alpha}$ is a constant. This
equation indicates that for the limits of $a^3\ll\alpha$, we have
$H(a)\approx H_0\sqrt{\frac{\alpha}{a^3}}$ leading to
$a(t)=a_{0m}t^{\frac{2}{3}}$ with the integration constant
$a_{0m}=(\frac{9}{4}H^{2}_{0}\alpha)^{1/3}$, which exactly is the
scale factor of the matter dominated era in the standard model of
cosmology. Moreover, for the limit of $a\gg1$, we have
$H(a)\rightarrow H_0$ leading to $a(t)=a_0\exp(H_0t)$ for the
scale factor of the current accelerating expansion phase of the
universe, in which $a_0$ is a constant.

Now, combining $1+z=a^{-1}$ with
$\lambda(a)=\frac{1}{4\kappa+\kappa\alpha a^{-3}}$ and inserting
the result into Eq.~(\ref{dece1}), we obtain
\begin{eqnarray}\label{dece2}
q(z)=\frac{\alpha(1+z)^3-2}{2(1+\alpha(1+z)^3)}.
\end{eqnarray}
In order to describe the evolution of the universe  from the
matter dominated era to the current accelerating phase, the
deceleration parameter $q(z)$ should satisfy the three
 conditions as ($i$)
$q(z\rightarrow\infty)\rightarrow\frac{1}{2}$, ($ii$)
$q(z\approx0.6)\rightarrow0$ and ($iii$)
$q(z\rightarrow0)\leq-\frac{1}{2}$ \cite{roos}. Using the equation
(\ref{dece2}), one can verify that the deceleration parameter of
the matter dominated era of the standard cosmology
($q=\frac{1}{2}$) is obtainable in the $\lambda\rightarrow0$ limit
or equivalently in the $\alpha\rightarrow\infty$ limit. Moreover,
at high redshift limit ($z\rightarrow\infty$) and independent of
the $\alpha$ parameter, we have $q\rightarrow\frac{1}{2}$ which
again addresses the matter dominated era. Therefore, the change of
the pressureless matter density in our model is the same as that
of the standard cosmology, i.e $\rho_m=\rho_0 a^{-3}$.
Additionally, although the deceleration parameter in our model
differs from that of the matter dominated era of the standard
cosmology, this era is covered at the appropriate limit of
$z\rightarrow\infty$ in our model. Here, from equation
(\ref{dece2}), for $-1<\alpha\leq\frac{1}{2}$, we have
$q(z=0)\leq-\frac{1}{2}$ which demonstrates the satisfaction of
the third condition. In addition, since there is no divergence  in
the history of the evolution of the universe from the early matter
dominated era to its current phase, $q(z)$ should not diverge
which requires that its denominator should not vanish for any
non-negative amount of $z$. This requires to have
$0\leq\alpha\leq\frac{1}{2}$ which consequently leads to the total
restricting range on the deceleration parameter as $-1\leq
q(z=0)\leq-\frac{1}{2}$.

\begin{figure}
\centering
\includegraphics[scale=0.4]{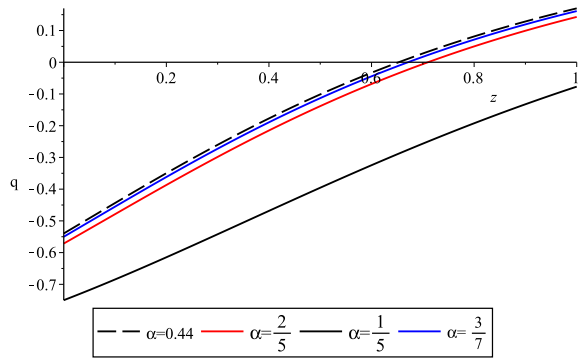}
\caption{Deceleration parameter $q$ versus the redshift $z$ for
some values of $\alpha$.} \label{fig}
\end{figure}

For example, consider the case of $q(z=0)=-0.55$ \cite{ref} which
through the equation (\ref{dece2}) corresponds to
$\alpha=\frac{3}{7}$. Considering this value of the $\alpha$
parameter, one can find that the $q=0$ case is associated to the
redshift $z\simeq0.67$ when the universe leaves its decelerating
phase and enters to the accelerating phase. This result is in
agreement with some observational evidences \cite{ref1,ref2,ref3}.
The deceleration parameter $q(z)$ is plotted in Fig.~(\ref{fig})
versus the redshift $z$ for some values of the $\alpha$ parameter.
It is seen from the figure that for the small redshifts,
representing the late time in the history of the universe, the
deceleration parameter goes to negative values representing an
accelerated expanding phase in our constructed model. As a result,
a non-minimal coupling between the geometry and pressureless
matter, which mainly consists of dark matter, may lead to a
description for the dark energy, and therefore the current
accelerating phase of the universe expansion.
\\

Based on the above results, this mutual relation between geometry
and the matter source suggests that this source and its enclosing
cosmic horizon may achieve the thermodynamic equilibrium, a result
which is in agreement with the recent study by Mimoso et al,
focusing on the properties and criterions of a thermodynamic
equilibrium between the cosmic horizon and the cosmic fluids in
various cosmic eras \cite{pavn}.

\section{Radiation dominated era and the curvature-radiation non-minimal coupling}
For the flat FRW universe filled by a radiation source, the
Friedmann equations are as follow
\begin{equation}\label{friedman11}
(12\kappa\lambda-3)H^2+6\kappa\lambda \dot{H}=-\kappa\rho_r,
\end{equation}
and
\begin{equation}\label{friedman21}
(12\kappa\lambda-3)H^2+(6\kappa\lambda-2) \dot{H}=
\frac{1}{3}\kappa\rho_r,
\end{equation}
where $\rho_r$ is the energy density. Because the energy-momentum
associated to radiation fields is a traceless source, i.e $T=0$,
by contracting equation~(\ref{gr}), one finds
$R(4\kappa\lambda-1)=\kappa T$ which clearly for
$\kappa\lambda\neq\frac{1}{4}$ results in a null Ricci scalar for
a radiation source, i.e $R=0$. Some simple calculations for the
continuity equation and deceleration parameter also lead to
\begin{eqnarray}\label{cont11}
\dot{\rho}_r&+&4H\rho_r =0,\\ \nonumber \rho_r&=&\rho_{0r}a^{-4},
\end{eqnarray}
where $\rho_{0r}$ is the integration constant, and
\begin{eqnarray}\label{dece11}
q(z)=1,
\end{eqnarray}
respectively. In order to obtain the last equation, we combined
Eqs.~(\ref{friedman11}) and~(\ref{friedman21}) with each other to
get $\frac{\dot{H}}{H^2}=-2$, a result which leads to
$a=a_0t^{\frac{1}{2}}$ for the scale factor where $a_0$ is the
integration constant, in agreement with the radiation dominated
era of the standard cosmology \cite{roos}. Based on
Eqs.~(\ref{cont11}) and~(\ref{dece11}), the density changes of the
radiation source and the deceleration parameter of the radiation
dominated era are the same as those of the standard cosmology
meaning that the radiation dominated era in our model is the same
as that of the standard cosmology. Indeed, since $R=0$ in the
radiation dominated era, independent of the value of $\lambda$
parameter we have $(\lambda R)^{;\nu}=0$ meaning that the above
results are independent of $\lambda$ parameter. Now, we use the
$\rho_m\rightarrow0$ limit of the $\lambda(a)$ relation obtained
in Eq.~(\ref{mym}), in order to find the value of $\lambda$ which
leads to $\lambda=\frac{1}{4\kappa}$. It means that since
$\lambda$ is a constant quantity, the geometry and radiation do
not affect each other.

Indeed, since radiation is a traceless source, simple calculations
lead to
\begin{eqnarray}\label{contn}
\dot{\rho}_r+4H\rho_r+\dot{\rho}_m+3H\rho_m=\frac{d}{dt}(\frac{\kappa\lambda}{4\kappa\lambda-1}\rho_m),
\end{eqnarray}
for the continuity equation in a universe filled by both radiation
and dust. In the absence of any interaction between radiation,
dust and geometry, this equation is decomposed to
Eqs.~(\ref{cont1}) and~(\ref{cont11}) meaning that the
$\rho_r=\rho_{0r}a^{-4}$, $\rho_m=\rho_0 a^{-3}$ and
$\lambda(\rho_m)=\frac{1}{4\kappa+\kappa C \rho_m}$ solutions are
also available in this case. Therefore, for $\rho_m=0$, we have
$\lambda=\frac{1}{4\kappa}$ meaning that the
$\lambda=\frac{1}{4\kappa}$ case is allowed in the radiation case.
Inserting $\lambda=\frac{1}{4\kappa}$ into either
Eqs.~(\ref{friedman11}) or~(\ref{friedman21}) and combining the
result with~(\ref{cont11}), we again reach at
$\frac{\dot{H}}{H^2}=-2$ leading to $a=a_0t^{\frac{1}{2}}$ and
thus $R=0$. Here, we should mention that since we have $R=T=0$ in
this era, the $R(4\kappa\lambda-1)=\kappa T$ condition is
available independent of the value of $\lambda$ parameter. Indeed,
unlike the Rastall theory, where the
$\lambda^{\prime}=\frac{1}{4\kappa}$ case is not allowed
\cite{rastall}, here, the $\lambda=\frac{1}{4\kappa}$ case can be
allowed. Therefore, although the geometry generally has the
ability to couple with the energy-momentum source in a non-minimal
way ($\lambda=constant\neq0$), since $\lambda$ is constant,
geometry and the radiation source do not affect each other. This
means that the ordinary energy-momentum conservation law is
respected by the radiation source as seen in~(\ref{cont11}).

Finally, we should mention that due to the fact that the radiation
source does not coupled to the geometry in a non-minimal way,
there is no energy flux between the geometry and radiation fields.
This may be considered as the reason for the failure to achieve
the thermodynamic equilibrium between the cosmic horizon and the
radiation fields \cite{pavn}.

\section{$\lambda$ and the primary inflationary era}

In this section, we address two methods to model the primary
inflationary era in our formalism, and also study the role and
behavior of $\lambda$ in these methods.

\subsection{$\lambda$ as the generator of the primary inflationary phase}

Now, let us consider an empty flat FRW universe with its
describing equations as
\begin{equation}\label{friedman110}
(12\kappa\lambda-3)H^2+6\kappa\lambda \dot{H}=0,
\end{equation}
and
\begin{equation}\label{friedman210}
(12\kappa\lambda-3)H^2+(6\kappa\lambda-2) \dot{H}=0.
\end{equation}
It is easy to check that both the above equations are true only
for $\lambda=\frac{1}{4\kappa}=constant$ and $\dot{H}=0$. It is
worthwhile mentioning that, as a desired result, the
$\lambda=\frac{1}{4\kappa}=constant$ solution is in full agreement
with the $\rho_m\rightarrow0$ limit of the results obtained in
Eqs.~(\ref{mym}) and~(\ref{contn}). Besides, since the spacetime
is empty ($T_{\mu\nu}=0$), we should have $\frac{d(\lambda
R)}{dt}=0$ meaning that $\lambda=\frac{\rho_g}{R}$, where
$\rho_g\equiv\psi$ is a constant. In addition, using the above
equations, one can obtain that the
$\lambda=\frac{1}{4\kappa}=constant$ and $\dot{H}=0$ conditions
lead to an exponential growth in the scale factor, i.e
$a(t)=a_0\exp(H_0t)$ where $a_0$ and $H_0$ are the integration
constants, with the non-vanishing Ricci scalar $R=12H_0^2$,
respectively. Now, combining the above results with each other, we
reach at $H_0=\sqrt{\frac{\kappa\psi}{3}}$. It is also obvious
that, since $\lambda$ and $\rho_g$ are constant, Eqs.~(\ref{gr0})
and~(\ref{emc}) are met here, and therefore, $\psi$ is nothing but
the integration constant in the RHS of Eq.~(\ref{emc1}). Indeed,
we should remind that, since Eq.~(\ref{emc1}) is the result of
Eq.~(\ref{gr0}) and thus Eq.~(\ref{emc}), the fulfillment of
Eq.~(\ref{gr0}) (or equally~(\ref{emc})) is necessary and
sufficient.

On the other hand, from Eq.~(\ref{gr}), we know that
$R(4\kappa\lambda-1)=T$ which its right hand side vanishes due to
the emptiness of the spacetime. Then, since the Ricci scalar does
not vanish, i.e $R\neq0$, we find out that we should have
$\lambda=\frac{1}{4\kappa}$. This is in agreement with the
previous mentioned results obtained from solving
Eqs.~(\ref{friedman110}) and~(\ref{friedman210}), and also
applying the $\rho_m\rightarrow0$ limit to Eq.~(\ref{mym}). Once
again, we see that unlike the original Rastall theory, the case of
$\lambda=\frac{1}{4\kappa}$ may be allowed in this new formulation
of the Rastall theory.

Therefore, the inflationary era may be supported in this model by
a unique feature of the geometry which is the ability of the
geometry to couple with the energy-momentum sources in a
non-minimal way in agreement with this fact that
$\lambda=\textmd{constant}\neq0$. In fact, the empty flat FRW
spacetime is forced to expand exponentially by this ability. We
should note that the absence of the energy-momentum source does
not mean that the geometry has not the ability of coupling to the
energy-momentum sources. Indeed, in this case, the absence of an
energy-momentum source only means that the geometry does not
couple to anything. It is also worthwhile to mention that since
$T=0$ and $\lambda$ is constant in both the radiation dominated
and the primary inflationary phases, the obtained results about
these eras may be generalizable to the original Rastall theory.

\subsection*{Energy extraction during the inflationary era}

We saw that the ability and tendency of geometry to couple with
the energy sources, in the non-minimal way, does not disappear,
i.e $\lambda\neq0$, in the absence of an energy-momentum source.
In fact, this is a property of geometry which enforces the empty
FRW spacetime to expand exponentially. Moreover, from
Eq.~(\ref{emc}), we found that the $\lambda R=\rho_g(\equiv\psi)$
term behaves as an energy density. Here, $\psi$ is the energy
density associated with the non-minimal coupling $\lambda$, and
therefore, we get $E=\int\psi dV=\frac{4\pi}{3}\psi a(t)^3V_0$ for
the total energy of co-moving volume $V_0$ corresponding to this
coupling at any given time $t$. Finally, for the amount of the
energy of the co-moving volume $V_0$ specified from spacetime at
time $t+\delta t$, due its intrinsic property to couple with the
energy-momentum sources in the non-minimal way, we have
\begin{eqnarray}
E(t+\delta t)=\frac{4\pi}{3}\psi
a(t+dt)^3V_0\simeq\frac{4\pi}{3}\psi
a(t)^3V_0\exp(3\sqrt{\frac{\kappa\psi}{3}}\delta t),
\end{eqnarray}
meaning that the released energy grows exponentially.

Therefore, in our formalism, the ability and tendency of geometry
to couple with the energy-momentum sources enforces universe to
expand, and in fact, it is the backbone of the universe expansion
and the energy production in the primary inflationary phase. Thus,
this ability  may also help us to provide a unified mechanism
explaining the primary inflationary era as well as the current
accelerating phase of the universe expansion.

\subsection{Standard inflation and $\lambda$}

In the previous subsection, we found out that, even in the absence
of an inflaton field, the tendency of geometry to couple with the
energy-momentum sources may lead to an inflationary phase for the
universe expansion, and consequently, the slow-rolling parameters
do not appear in that scenario. It is useful to mention that there
are also some inflationary models in which the slow-roll condition
does not appear \cite{g1,g2,g3}. Here, we will show that the
standard inflation scenario by implementing an inflaton field can
also be valid in our formalism.

In order to achieve this goal, we consider a spatially homogeneous
scalar field evolving in potential $\mathcal{V}(\phi)$. Therefore,
simple calculation yields
$\rho_\phi=\frac{1}{2}\dot{\phi}^2+\mathcal{V}(\phi)$ and
$p_\phi=\frac{1}{2}\dot{\phi}^2-\mathcal{V}(\phi)$ for the energy
density and pressure of the inflaton field \cite{roos}. Now, the
Friedmann equations in a flat FRW universe filled by the mentioned
field are written as
\begin{eqnarray}\label{fr01}
(12\kappa\lambda-3)H^2+6\kappa\lambda \dot{H}&=&-\kappa\rho_\phi,\nonumber\\
(12\kappa\lambda-3)H^2+(6\kappa\lambda-2) \dot{H}&=&\kappa p_\phi,
\end{eqnarray}
which finally lead to
\begin{eqnarray}\label{fr001}
\dot{H}&=&-\frac{\kappa}{2}[\rho_\phi+p_\phi],
\end{eqnarray}
for the Raychaudhuri equation. In addition, the same as the matter
dominated era, considering a simple situation in which there is no
energy exchange between the energy-momentum source and geometry,
we reach at $\dot{\rho}_\phi+3H(\rho_\phi+p_\phi)=-\frac{d(\lambda
R)}{dt}=0$ leading to
\begin{eqnarray}\label{my}
&&\ddot{\phi}+\frac{\partial
\mathcal{V}}{\partial\phi}+3H\dot{\phi}=0,\nonumber\\
&&\lambda=\frac{1}{4\kappa[1+\gamma^{-1}(3p_\phi-\rho_\phi)]},
\end{eqnarray}
for the continuity equation in which $\gamma^{-1}$ is constant.
Now, if $\ddot{\phi}$ is negligible and
$\dot{\phi}^2\ll\mathcal{V}(\phi)$, then $p_\phi\simeq-\rho_\phi$
and from Eqs.~(\ref{fr001}) and~(\ref{my}), we find that
$\dot{H}\simeq0$ and
$\lambda(\phi)\simeq\frac{1}{4\kappa[1-4\gamma^{-1}\mathcal{V}(\phi)]}$,
respectively. In fact, when $\ddot{\phi}$ is negligible,
Eq.~(\ref{my}) helps us in getting
$\mathcal{V}^{\prime\prime}\simeq\frac{3\kappa}{2}\dot{\phi}^2$
leading to
$\eta\equiv\frac{2}{3\kappa}(\frac{\mathcal{V}^{\prime\prime}}{\mathcal{V}})\simeq\frac{\dot{\phi}^2}{\mathcal{V}}$.
Therefore, during the inflation process, when the slow-roll
approximation is valid, we have $\eta\ll1$ in agreement with the
standard inflation hypothesis \cite{roos}. In this manner,
inserting Eq.~(\ref{my}) into Eq.~(\ref{fr01}), one can easily
reach at $H^2\simeq\frac{\kappa}{3}[4\mathcal{V}-\gamma]$
recovering the standard inflation results at the appropriate limit
of $\lambda\rightarrow0$ (or equally $\gamma\rightarrow0$).
Moreover, since $q=-1-\frac{\dot{H}}{H^2}\simeq-1$ at the time of
inflation, we should have $\epsilon\equiv-\frac{\dot{H}}{H^2}\ll1$
\cite{roos}. Now, using Eqs.~(\ref{fr001}) and~(\ref{my}), one
obtains
$\epsilon\simeq\frac{8}{\kappa}[\frac{\mathcal{V}^\prime}{4\mathcal{V}-\gamma}]^2$,
where the prime sign stands for the derivative with respect to
$\phi$. It is interesting to note that if we define
$\tilde{V}(\phi)\equiv4\mathcal{V}-\gamma$, then we have
$H^2\simeq\frac{\kappa}{3}\tilde{V}$ and
$\epsilon\simeq\frac{8}{\kappa}[\frac{\mathcal{V}^\prime}{4\mathcal{V}-\gamma}]^2=\frac{1}{2\kappa}(\frac{\frac{\partial\tilde{V}}{\partial\phi}}{\tilde{V}})^2$
similar to those of the standard inflation scenario \cite{roos}.
Therefore, if the slow-roll approximation is valid, then a
spatially homogeneous scalar field evolving in potential
$\mathcal{V}(\phi)$ can support the primary inflationary era in
our formalism whenever the inflaton field (or equally
$\mathcal{V}(\phi)$) satisfies the $H^2\simeq constant>0$,
$\epsilon<1$ and $\eta\ll1$ conditions. It is finally worth to
mention that approaching the end of inflation, where
$\mathcal{V}(\phi)\rightarrow0$, we have
$\lambda\rightarrow\frac{1}{4\kappa}$ revealing the consistency
with our results in previous sections about the radiation
dominated era.

\section{Summary and Concluding Remarks}

After referring to the Rastall theory, we addressed a
generalization of this theory and studied some of its cosmological
consequences. Based on our results, a non-minimal coupling between
the geometry and a pressureless matter field may lead to a
transition from the matter dominated era to the current
accelerating phase, in agreement with some previous observations
\cite{ref1,ref2,ref3}. We only focused on the $T^{\mu \nu}_{\ \
;\mu}=0=-\frac{d(\lambda R)}{dt}$ solutions. In this case, a dust
source, which satisfies the ordinary energy-momentum conservation
law, is allowed, and as we have seen, the evolution of its energy
density is the same as that of the standard cosmology. It should
also be noted that although the same as the general relativity
$T^{\mu \nu}_{\ \ ;\mu}=0$ in our model, since $\lambda\neq0$,
Friedmann equations in our model differ from those of the standard
cosmology. In addition, we found out that, in our formalism, the
evolution of the energy density in the radiation dominated era is
the same as that of the standard cosmology. Indeed, we found that,
during the radiation dominated era, $\lambda$ remains a non-zero
constant quantity meaning that the evolution of the radiation
source as well as the geometry do not affect the value of
$\lambda$.

Finally, we considered an empty flat FRW universe and realized
that, even in the absence of an inflaton field, a primary
inflationary era can be driven in this generalized version of
Rastall theory when $\lambda=\frac{1}{4\kappa}$. Therefore, our
study shows that the ability and tendency of geometry to couple
with the energy-momentum sources ($\lambda\neq0$) may be the
backbone of the primary inflationary era and the current
accelerating phases of the universe expansion in a unified
picture. Also, a scenario for a universe filled by an inflaton
field in the context of the Rastall theory has been introduced. In
this context, as the matter dominated era, we have only focused on
simple case of $T^{\mu \nu}_{\ \ ;\mu}=0=-\frac{d(\lambda R)}{dt}$
meaning that there is no energy exchange between geometry and the
cosmic fluid. Once again, we should remind that since
$\lambda\neq0$, the Friedmann equations in our model differ from
those of the general relativity. It is obtained that if the
inflaton field meets the usual slow-roll conditions, then it can
support an inflationary phase.

\section*{Acknowledgment}
We are so grateful to the respected referees for their valuable
comments. The work of H. Moradpour has been supported financially
by Research Institute for Astronomy \& Astrophysics of Maragha
(RIAAM).

\end{document}